\documentclass[11pt]{article}
\usepackage{moriond,epsfig,psfig}

\def\be{\begin{equation}}
\def\ee{\end{equation}}
\def\bea{\begin{eqnarray}}
\def\eea{\end{eqnarray}}

\def\chandra{{\em Chandra}}

\begin{document}
\vspace*{4cm}
\title{Particle Acceleration in Supernova Remnants and Pulsar Wind Nebulae}

\author{Patrick Slane}

\address{Harvard-Smithsonian Center for Astrophysics\\ 60 Garden Street\\
Cambridge, MA 02138, USA}

\maketitle\abstracts{
While supernova remnants (SNRs) have long been considered prime candidates for
the source of cosmic rays, at least to energies up to $\sim 10^{14}$~eV, 
it is only over the past several years that direct evidence of such energetic
particles in SNRs has been uncovered. X-ray observations of several shell-type
SNRs have now revealed sites dominated by nonthermal emission, indicating
an electron population whose energy extends far beyond the thermal
distribution typical of such SNRs. In other remnants, discrepancies between
the shock velocity and the electron temperature points to a strong cosmic
ray component that has essentially thrived at the expense of the thermal
component of the gas. Modeling of the radio, X-ray, and gamma-ray emission
provides strong constraints on the acceleration mechanism as well as the
properties of the ambient medium in which the mechanism prospers.
In the innermost regions of some SNRs, particle acceleration is taking place
over much different scales. The formation of Crab-like pulsar wind nebulae
(PWNe) is understood to require the presence of a termination shock at 
which the relativistic pulsar wind is forced to join the slow expansion 
of the outer nebula. While the acceleration mechanism is necessarily 
different, these
shocks also act as sites in which particles are boosted to high energies.
In the Crab Nebula, optical wisps mark the location of this termination shock.
Recent X-ray observations have begun to reveal the termination shock zones
in other PWNe, and are now allowing us to constrain the nature of the pulsar
wind as well as the flow conditions in the outer nebula. Here I 
present a summary of the properties of shock acceleration in these two 
distinct regions of SNRs, and review recent observational results in which 
the properties of the shocks are finally being revealed. 
}

\section{Introduction}

Supernova remnants and their associated pulsars 
are sites in which strong shocks act to accelerate
particles to extremely high energies. The connection between SNRs and 
the energetic cosmic rays that pervade the Galaxy has long been assumed, 
for example; shock acceleration by the SNR blast wave provides ample 
energy for the production of multi-TeV particles, and recent observations
of nonthermal X-ray and VHE $\gamma$-ray emission from several
SNRs has confirmed the presence of electrons at these high energies.
At the same time, models for the structure of PWNe predict particle 
acceleration at the wind termination shock, and recent X-ray observations
have begun to probe these acceleration sites.  Here I review recent
and ongoing observational X-ray studies of particle acceleration by shocks
in SNRs and PWNe.

\section{Shocks and Particle Acceleration in SNRs}

The birthrate and overall energetics of SNRs provide a strong plausibility
argument that they are a major source of cosmic rays up to the ``knee''
of the spectrum. 
However, until recently the only direct evidence of energetic particles
from SNRs has been from the radio synchrotron emission. But this corresponds
to electron energies 
\begin{equation}
E_{\rm GeV} \approx \left[\frac{\nu}{16 {\rm\ MHz}} B_\mu^{-1}\right]^{1/2}
\end{equation}
where $\nu$ is the frequency of the radio emission
and $B_\mu$ is the magnetic field strength in microGauss. Thus, while providing
a clue, the radio emission samples only electrons far below the cosmic ray 
knee.
X-ray observations allow us to characterize the thermodynamic states of 
SNRs and to infer the dynamics of their evolution, probing much higher 
particle energies.
As the blast wave from a supernova explosion expands, material from
the surrounding circumstellar material (CSM) and ISM is swept up
into a shell of hot gas. For an ideal gas, the
shock jump conditions yield an increase in density by a factor of
4 and a postshock temperature of
\begin{equation}
T = \frac{3 \mu m}{16 k} V_s^2
\end{equation}
where $V_s$ is the shock speed, $m$ is the proton mass, and $\mu$ is
the mean molecular weight of the gas ($\mu \approx 0.6$).
This shock-heated gas yields the familiar X-ray emission, characterized
by a thermal bremsstrahlung spectrum accompanied by strong emission lines.
As increasing amounts of material are swept up by the blast wave, the
shock is decelerated and the ejecta from the explosion encounter the
dense shell. A reverse-shock is generated, heating the ejecta. At early
times, then, the X-ray spectrum is dominated by emission from the ejecta.
As the amount of swept-up material increases, the spectrum becomes
dominated by emission from material with ISM abundances.

In addition to thermal heating of the swept-up gas, some fraction of the
shock energy density goes into nonthermal production of
relativistic particles through diffusive shock acceleration.
The maximum particle energy in such a scenario can be limited by radiative
losses, the finite age of the SNR, or particle escape from the accelerating
region. Radio observations
of SNRs provide ample evidence of electrons with GeV energies through
synchrotron radiation in the compressed magnetic field at the remnant
shell. At higher particle energies, $\gamma$-ray production may
result from nonthermal bremsstrahlung of electrons colliding with
ambient gas, from inverse Compton scattering of ambient photons,
and from the decay of neutral pions created by the collision of
energetic protons. If the relativistic particle component of the
energy density becomes comparable to that of the thermal component,
the shock acceleration process can become highly nonlinear. The gas
becomes more compressible, which results in a higher density and
enhanced acceleration. 
When the acceleration is very efficient, the relationship between the
shock velocity and the mean postshock temperature is no longer well
approximated by Eq. 1; the acceleration process depletes 
thermal energy, and the temperature for a given shock velocity drops below
that expected in the test particle case~\cite{anne}.
This process has been considered in detail
by Baring et al.~\cite{baring} who present a model for the radio to
$\gamma$-ray emission. 
The broadband spectra of SNRs depend highly on ambient conditions,
and X-ray studies of these SNRs reveal these conditions and can provide
spectral measurements which strongly constrain the models. 

In the simplest picture, the passage of material through the SNR shock results
in electrons and ions being boosted to the velocity of the shock. Because of
the mass difference, this means that the electrons and ions are not initially
in temperature equilibrium. The maximum timescale for equilibration is that
provided by Coulomb interactions, but plasma processes may reduce this
considerably. The state of equilibration is important, because while
the dynamics of SNR evolution are dominated by the ions (which carry the bulk
of the momentum), it is the electrons that produce the X-ray emission we
observe. Thus, when temperature measurements are used to infer the shock
velocity, for example, the state of temperature equilibration is exceedingly
important. \chandra\ HETG observations of SN~1987A,~\cite{michael} 
for example, yield spectra consistent with an electron 
temperature of 2.6~keV. However, the broadened line profiles indicate a 
blast wave speed of $\sim 3500 {\rm\ km\ s}^{-1}$ which corresponds to 
a post-shock temperature of 17~keV, providing evidence for incomplete 
electron-ion temperature equilibration. Similarly, the blast wave 
speed inferred from an expansion
study comparing a \chandra\ image of 1E~0102.2--7219 with high resolution images
taken with {\it Einstein} and {\it ROSAT} indicates a post-shock temperature
which is much higher than the observed electron temperature.~\cite{hughes} 
In this case, the discrepancy appears
to be larger than can be accounted for assuming Coulomb equilibration,
suggesting that a significant fraction of the shock energy has gone into
cosmic ray acceleration rather than thermal heating of the postshock gas.
Higher resolution X-ray expansion studies of 1E~0102.2--7219 (and other
young SNRs) are needed to
confirm this scenario, but the notion that particle acceleration is
efficient enough in some SNRs to significantly affect their dynamics
has strong foundations from recent studies of other SNRs. As we discuss
below, direct evidence of very energetic electrons now exists for a
handful of shell-type SNRs. In addition to the three SNRs which we
discuss in the following sections, for which synchrotron radiation dominates 
the X-ray emission,
evidence for energetic particles in the form of hard
tails in the X-ray spectra have also been observed for Cas~A~\cite{allen97,vink}
RCW~86,~\cite{allen99,bork} and other SNRs.

\subsection{SN~1006}
The first evidence of multi-TeV particles directly associated with 
a shell-type SNR was uncovered in studies of SN~1006 with the ASCA 
observatory.~\cite{koyama95} While spectra from the central 
regions of the SNR show distinct line emission associated with shock-heated
gas, the emission from the bright limbs of the remnant was shown to be
completely dominated by synchrotron emission. Reynolds~\cite{reynolds} modeled 
the emission as the result of diffusive shock acceleration in a low
density medium with the magnetic field orientation providing the distinct
``bilateral'' morphology. Subsequent observations with the CANGAROO 
telescope revealed VHE $\gamma$-ray emission from one limb of the 
SNR,~\cite{tanimori}
thus confirming the presence of extremely energetic particles. The low
ambient density of this remnant, which resides well above the Galactic
Plane, is insufficient to explain the $\gamma$-ray emission as the result
of $\pi^0$ decay from proton-proton collisions. Rather, the emission
results from inverse-Compton scattering of microwave background photons
off the energetic electron population in SN~1006. Using joint spectral
fits to the radio as well as thermal and nonthermal X-ray emission, 
Dyer et al.~\cite{dyer} conclude that the total energy in relativistic
particles is $\sim 100$ times the energy in the magnetic field, confirming
the notion that the nonthermal particle component contributes 
significantly to the dynamics of the blast wave evolution.

\subsection{G347.3--0.5 (RX J1713.7-3946)}

\begin{figure}[tb]
\centerline{\psfig{file=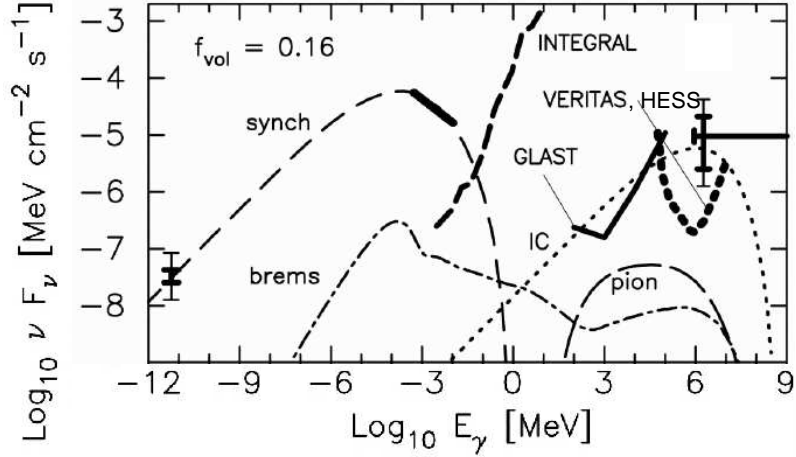,height=2.5in}}
\caption{
Comparison of broadband emission from the northwest limb of G347.3--0.5
with diffusive shock acceleration model results (from Ellison, Slane,
\& Gaensler 2001). The radio and X-ray emission is from synchrotron radiation
while the VHE $\gamma$-ray emission is from inverse-Compton scattering.
Sensitivities of future high energy telescopes are shown for comparison.
}
\end{figure}

ASCA observations of G347.3--0.5~\cite{koyama97,slane99} 
established this as the second member of the class of shell-type SNRs for
which the X-ray flux is dominated by synchrotron radiation. Unlike 
SN~1006, this SNR appears to have evolved in the vicinity of dense
molecular clouds~\cite{slane99} with which the shock may now be
interacting. CANGAROO observations~\cite{muraishi}
reveal VHE $\gamma$-ray emission from 
the vicinity of the northwest rim, which is brightest in X-rays. Combining
radio measurements from the ATCA with the X-ray and $\gamma$-ray 
results, Ellison, Slane, \& Gaensler~\cite{ellisonetal}
used diffusive shock acceleration models to
conclude that the radio and X-ray emission results from synchrotron radiation 
from a nonthermal electron population accelerated by the forward shock, and 
that the $\gamma$-ray
emission can be self-consistently modeled as inverse-Compton emission
(Figure 1).
Combined with limits on the ambient density based on the lack of thermal
X-ray emission, the models indicate very efficient particle acceleration
with $> 25\%$ of the shock kinetic energy going into relativistic 
ions. A comparison of Chandra observations of the northwest rim with
high resolution radio maps from the ATCA 
shows good overall agreement with the radio and X-ray morphology in this 
region (Figure 2), consistent with the interpretation that the emission
comes from the same electron population.~\cite{lazendic}

The nearby unidentified EGRET source 3EG J1714-3857 has been suggested as
being associated with G347.3--0.5, possibly 
resulting from the decay of neutral pions produced in the collision
of accelerated ions with the nearby molecular clouds,~\cite{butt} although this
would appear inconsistent with the X-ray measurements unless the emission
originates from a distinct spatial region. Most recently, Enomoto et 
al.~\cite{enomoto}
present new CANGAROO data which, they argue, establishes $\pi^0$-decay
as the mechanism by which the TeV $\gamma$-rays are produced. However,
the predicted spectrum overpredicts emission from the EGRET band by a
large margin,~\cite{reimer} making the claim for direct observation
of ion acceleration appear problematic.

\begin{figure}[h]
\centerline{\psfig{file=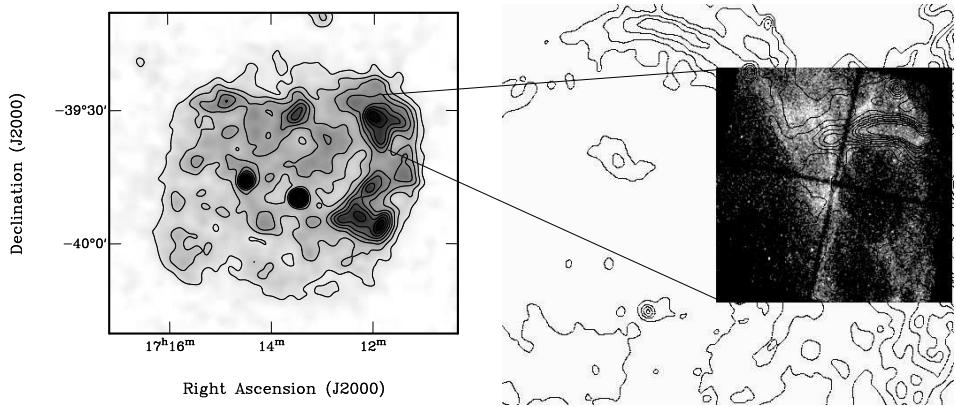,height=2.1in}}
\caption{
Left: ROSAT PSPC image of G347.3$-$0.5. Right: {\em Chandra}
image of northwest rim of G347.3$-$0.5, with radio contours from ATCA.
}
\end{figure}

\subsection{G266.2--1.2 (RX J0852.0-4622)}

G266.2--1.2 is an SNR lying along the line of sight to the southeast
corner of the Vela SNR.
Discovered in the ROSAT All-Sky Survey,~\cite{aschenbach} the remnant
is evident only at higher X-ray energies where the very soft spectrum of
Vela does not contribute. ASCA observations~\cite{slane2001} show that
the X-ray emission from this remnant is also predominantly synchrotron
radiation. To date, no high energy $\gamma$-ray emission has been reported
for this remnant, although there have been claims of $^{44}$Ti line
emission~\cite{iyudin} which, given the very short half-life, 
would require an extremely young age. The radio flux from G266.2--1.2
is extremely low,~\cite{duncan} much like that of SN~1006 and
G347.3--0.5. The overall properties of this remnant are indicative of
efficient shock acceleration of particles to very high energies, but
further investigations are required to quantify the dynamics.

\section{Energetic Particles from Pulsars and Their Wind Nebulae}

Our basic understanding of ``Crab-like'' SNRs stems from the picture presented
by Rees \& Gunn~\cite{rees} , and expanded upon by Kennel \& 
Coroniti,~\cite{kennela,kennelb}
in which an energetic wind is injected from a
pulsar into its surroundings. 
As illustrated schematically in Figure~3,
the basic structure of a pulsar wind nebula is regulated by the input
power from the pulsar and the density of the medium into which the nebula
expands; the pulsar wind inflates a magnetic bubble which
is confined in the outer regions by the expanding shell of ejecta or
interstellar material swept up by the SNR blast wave. The boundary condition
established by the expansion at the nebula radius $r_N$ results in the
formation of a wind termination shock at which the highly relativistic
pulsar wind is decelerated to $v \approx c/3$ in the postshock region,
ultimately merging with the particle flow in the nebula. The
shock forms at the radius $r_w$ at which the ram pressure of the wind is
balanced by the internal pressure of the pulsar wind nebula:
\begin{equation}
r_w^2 = \dot E/(4 \pi \eta c p)
\end{equation}
where $\dot E$ is the rate at which the
pulsar injects energy into the wind, $\eta$ is the fraction of a spherical
surface covered by the wind, and $p$ is the total pressure outside the shock.

\begin{figure}[h]
\centerline{\psfig{file=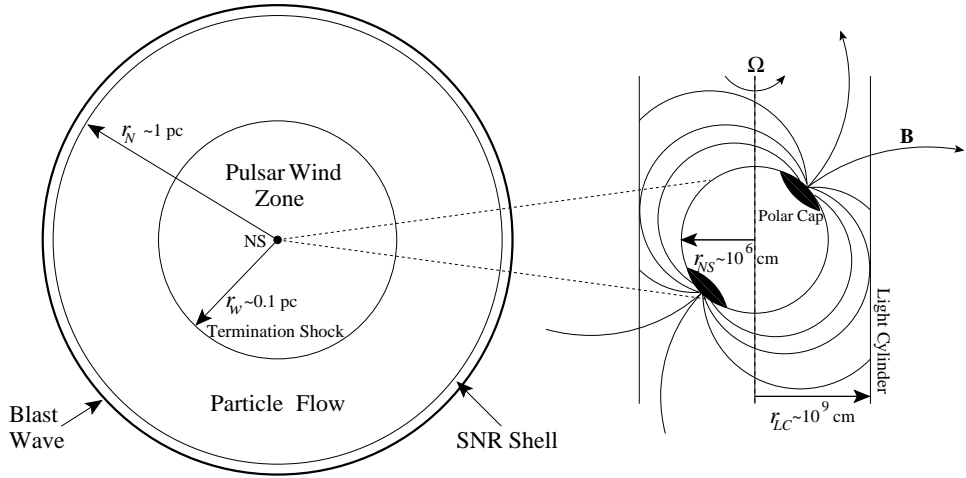,width=5in}}
\caption{Schematic view of a pulsar and its wind nebula. See the text for
a complete description. (Note the logarithmic
size scaling in the PWN figure when comparing with images shown elsewhere in
the text.)}
\end{figure}

The dynamics of the particle flow yield $\gamma \sim 10^6$ for the electrons
in the postshock region.~\cite{arons_tavani}
However, for typical magnetic
field strengths the observed X-ray emission requires $\gamma > 10^8$.
Particle acceleration at the termination shock apparently boosts the
energies of the wind particles by a factor of 100 or more, to energies
in excess of $\sim 50$~TeV. Arons \& Tavani~\cite{arons_tavani} (see 
also Arons~\cite{arons})
note that this process cannot proceed by normal diffusive shock acceleration
because the magnetic field at the termination shock must be nearly
perpendicular to the flow. Rather, they argue that the e$^\pm$ acceleration
is the result of resonant cyclotron absorption of low frequency electromagnetic
waves emitted by ions gyrating in the compressed $B$-field of the hot
post-shock gas. 
Ultimately, the pressure in the nebula is believed to reach the equipartition
value; a reasonable pressure estimate can be obtained by integrating
the radio spectrum of the nebula, using standard synchrotron emission
expressions, and assuming equipartition between particles and the magnetic
field. Typical values yield termination shock radii of order 0.1~pc, which
yields an angular size of several arcsec at distances of a few kpc. 

The magnetic field in the PWN builds up with
radius as the result of particle currents and wound-up magnetic flux from
beyond the pulsar light cylinder at $r_{LC}=c/\Omega$.
The finite lifetime of the synchrotron-emitting particles in the nebula,
$$t_{syn} \approx 5 \times 10^{11}/(B^{3/2} \nu^{1/2}) {\rm\ s}$$
(where $\nu$ is the frequency of the synchrotron radiation),
results in fewer high energy
particles at large radii. This manifests itself as a variation of the nebula
radius with energy (which is clearly seen in the Crab Nebula, for which the
radio size is larger than that seen in X-rays) or, equivalently, as a
radial variation in the X-ray spectrum, with the power law index steepening
with
radius.

The pulsar wind can be characterized in terms of $\dot E$
and the parameter parameter $\sigma$ representing
the ratio of Poynting flux to particle flux. For the Crab Nebula,
Kennel \& Coroniti~\cite{kennela} find
that small values of $\sigma$ ($\sim 10^{-3}$) are required, indicating a
particle-dominated wind. Yet current understanding of pulsar outflows
predicts $\sigma \gg 1$ where the wind is launched.~\cite{arons} Somehow,
between the pulsar light cylinder and the wind termination shock the 
balance between the electromagnetic energy and the kinetic energy of the
flow changes dramatically. The ability to identify the termination shock
and measure the emission parameters in the immediate region is now 
providing constraints on the nature of the wind that may ultimately
help unravel this problem. Below I summarize recent X-ray investigations
that have finally begun to probe this important shock region in which
particle acceleration is apparently taking place.

\begin{figure}[t]
\centerline{\psfig{file=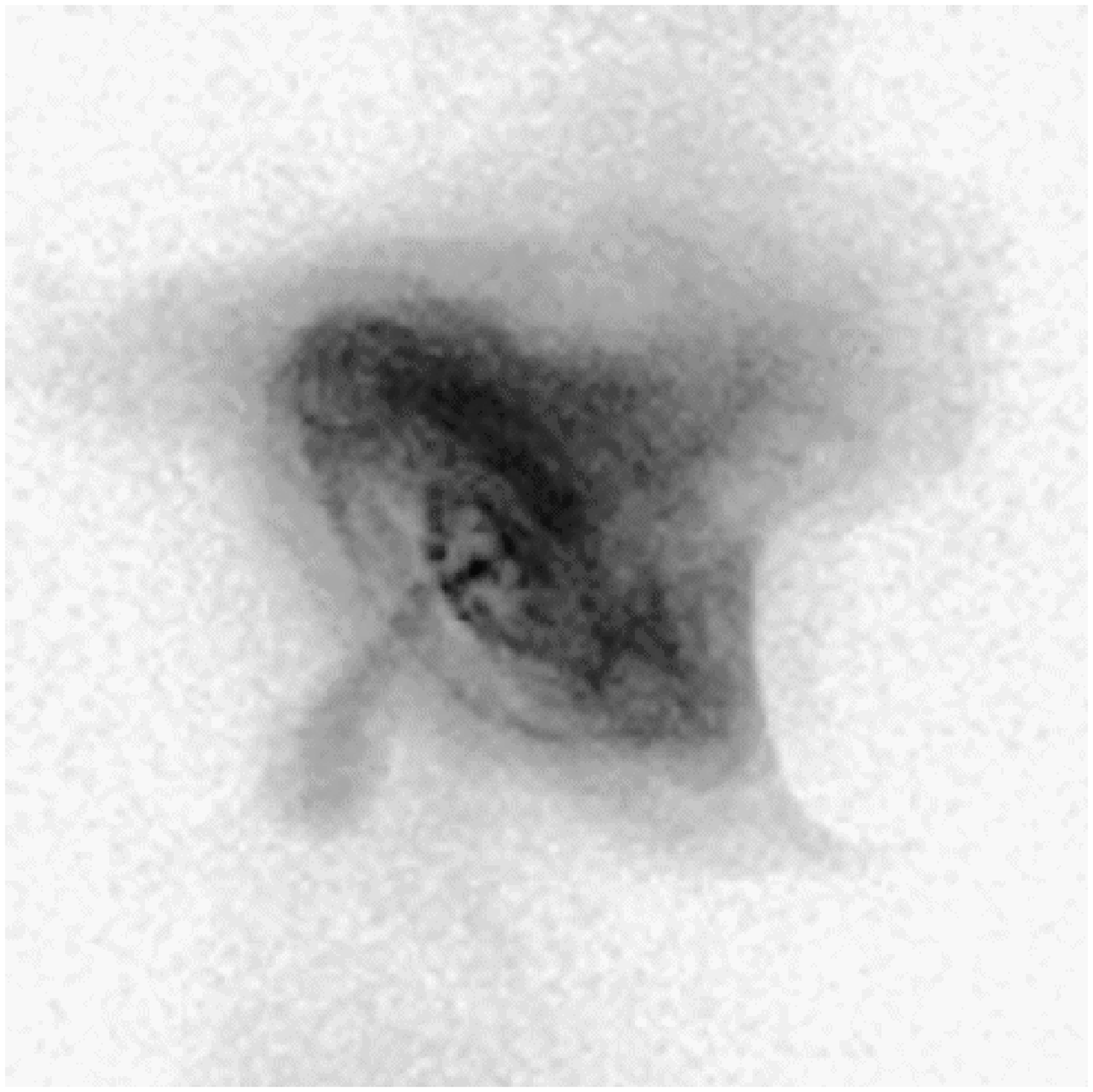,height=2.3in}
\psfig{file=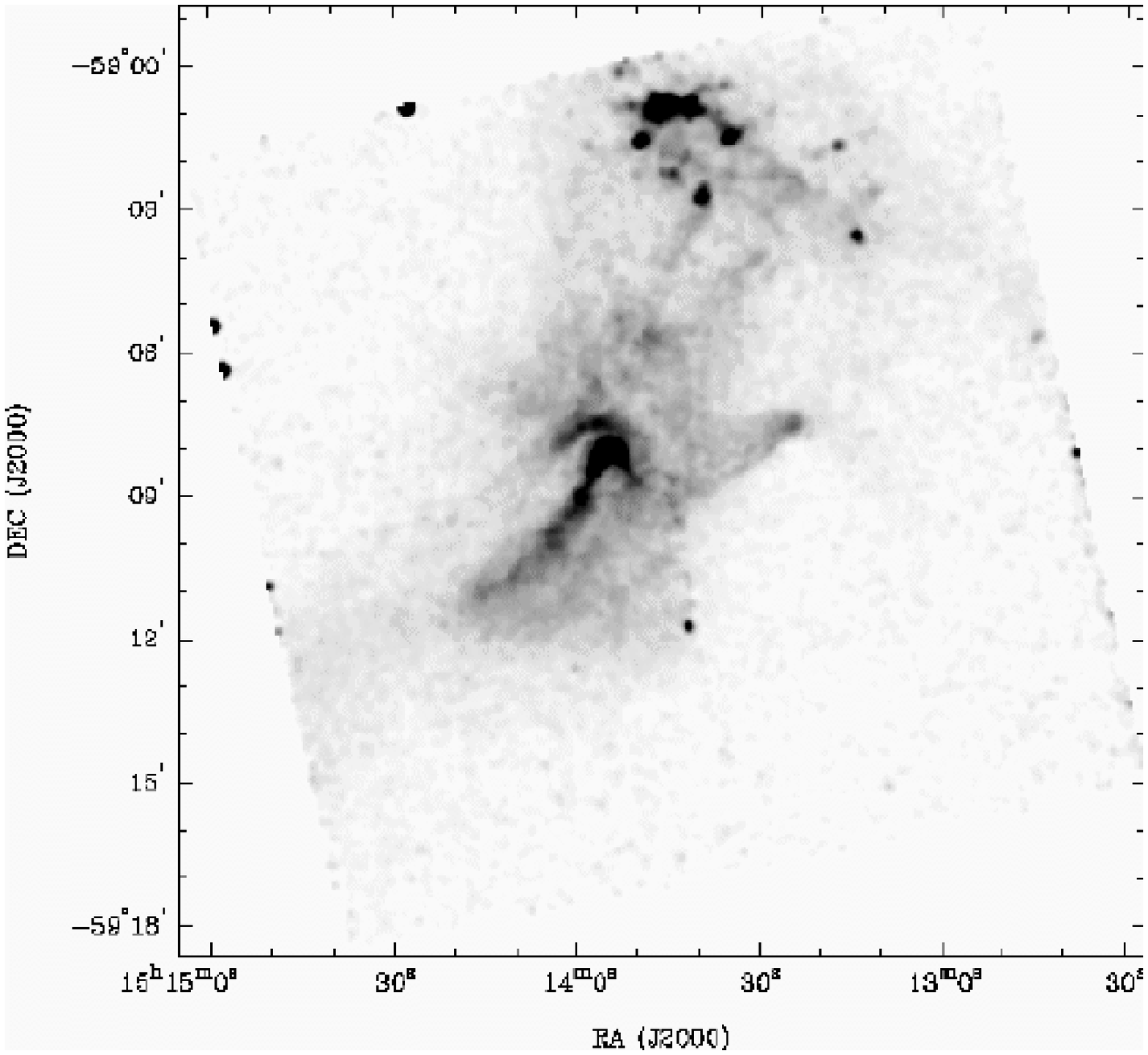,height=2.3in}}
\caption{
{\em Chandra} image of the Crab Nebula (left) and the pulsar wind
nebula associated with PSR~B1509--58 (right; image courtesy of B. Gaensler).
}
\end{figure}

\subsection{Crab Nebula}

The Crab Nebula is the best known of the class of pulsar wind nebulae 
and has inspired much of the theoretical work on PWNe. Powered
by an energetic central pulsar, it emits synchrotron radiation from radio
wavelengths up beyond the hard X-ray band. Optical wisps are observed
in the inner nebula, at a position interpreted as the pulsar wind termination
shock,~\cite{hester95} and high resolution X-ray observations 
(Figure 4, left) reveal a distinct ring of emission in this same region 
as well as a jet emanating from the pulsar.~\cite{weisskopf}
Moreover, monitoring observations of the Nebula~\cite{hester95,mori} 
show that these and other detailed features are highly dynamic. 
The discovery of radio wisps in inner ring region~\cite{bietenholz}
suggests that the acceleration site may be the same for the entire population
of electrons that produce the broad-band synchrotron emission. 

In addition to the jet and inner ring, the X-ray image reveals an outer
toroidal structure that presumably lies in the equatorial plane, as well as
fine structure correlated with optical polarization measurements, indicating 
that the structures trace the magnetic field. The early models
of Rees \& Gunn~\cite{rees}  and Kennel \& Coroniti~\cite{kennela,kennelb}
predict these basic
properties as the result of a wound-up magnetic field, the large-scale
confinement of the Nebula by the (unseen) supernova ejecta, and the termination
of the pulsar wind flow by an inner shock. This picture leads to the
inference of a
low-$\sigma$ wind described above. As we describe below, recent observations
of PWNe have begun to show many similar features, indicating that our basic
picture -- while still poorly understood in detail -- can at least be said
to apply to a ``class'' of objects.

\subsection{PSR B1509--58}
The PWN powered by the energetic young pulsar B1509--58 (Figure 4, right)
displays a complex morphology rivaling that of the Crab Nebula. \chandra\
observations~\cite{gaensler} reveal a clear symmetry axis, presumably
corresponding to the pulsar spin axis, with an apparent curved jet
along this axis. Arc-like features in the inner nebula appear to 
correspond to equatorial structures similar in nature and origin to
the wisps seen inside the Crab X-ray torus. Gaensler et al.~\cite{gaensler} 
show that $\sigma \sim 0.005$ at these wisps, with 
other compact knots seen very close to the pulsar indicating even
smaller $\sigma$ values upstream of the wind shock zone.
The PWN has a much weaker magnetic field than the Crab, and synchrotron
losses affect the spectra of outflowing particles only at large radii.
The hard spectra observed
for the compact knots suggest a different acceleration mechanism than
for larger features, downstream of the termination shock. 
In the northernmost regions of the PWN, clumps of thermally-emitting gas
from the associated SNR appear to be embedded in the nebula.

\subsection{G21.5--0.9}

\begin{figure}[tb]
\centerline{\psfig{file=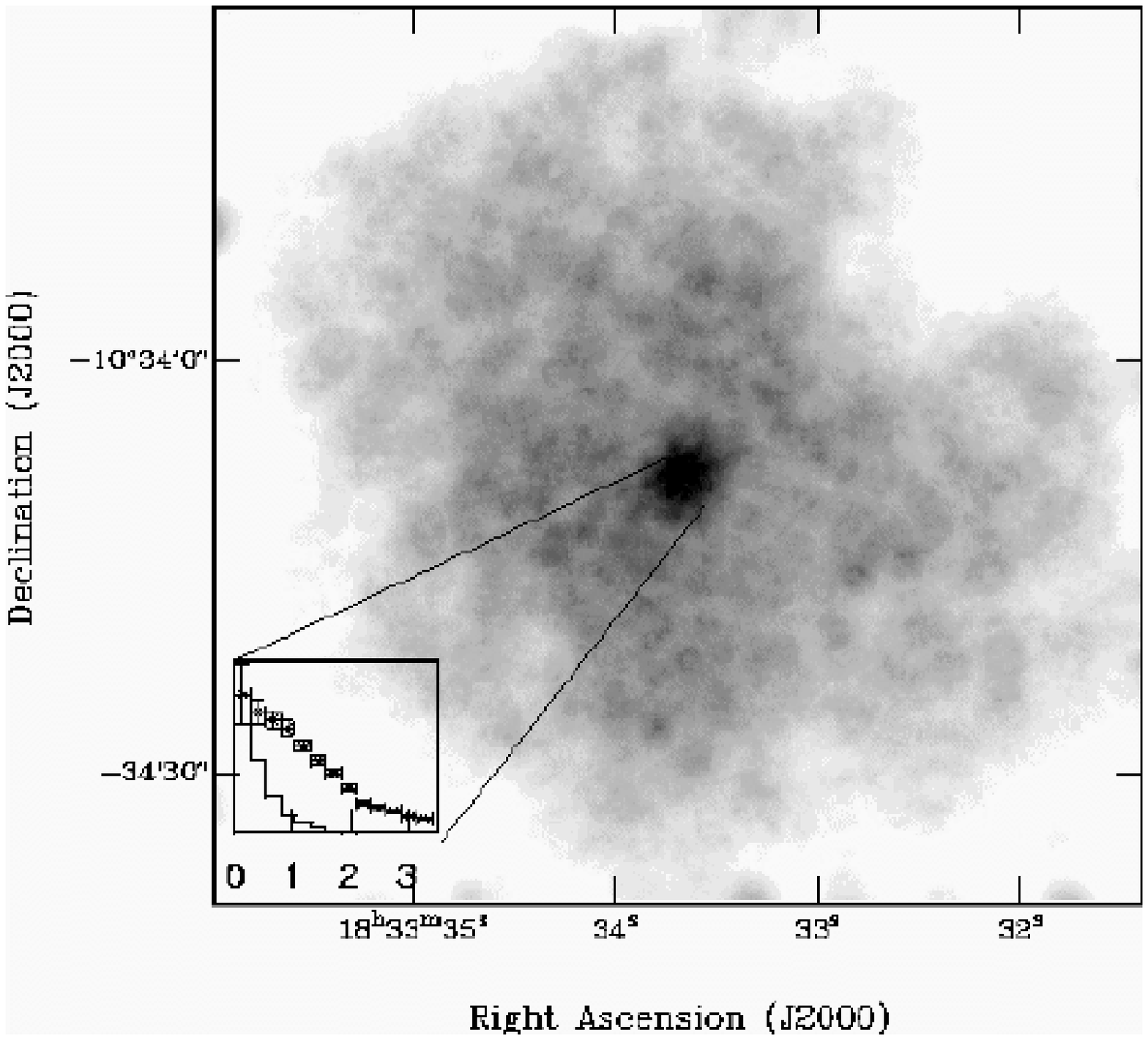,height=2.5in}
\psfig{file=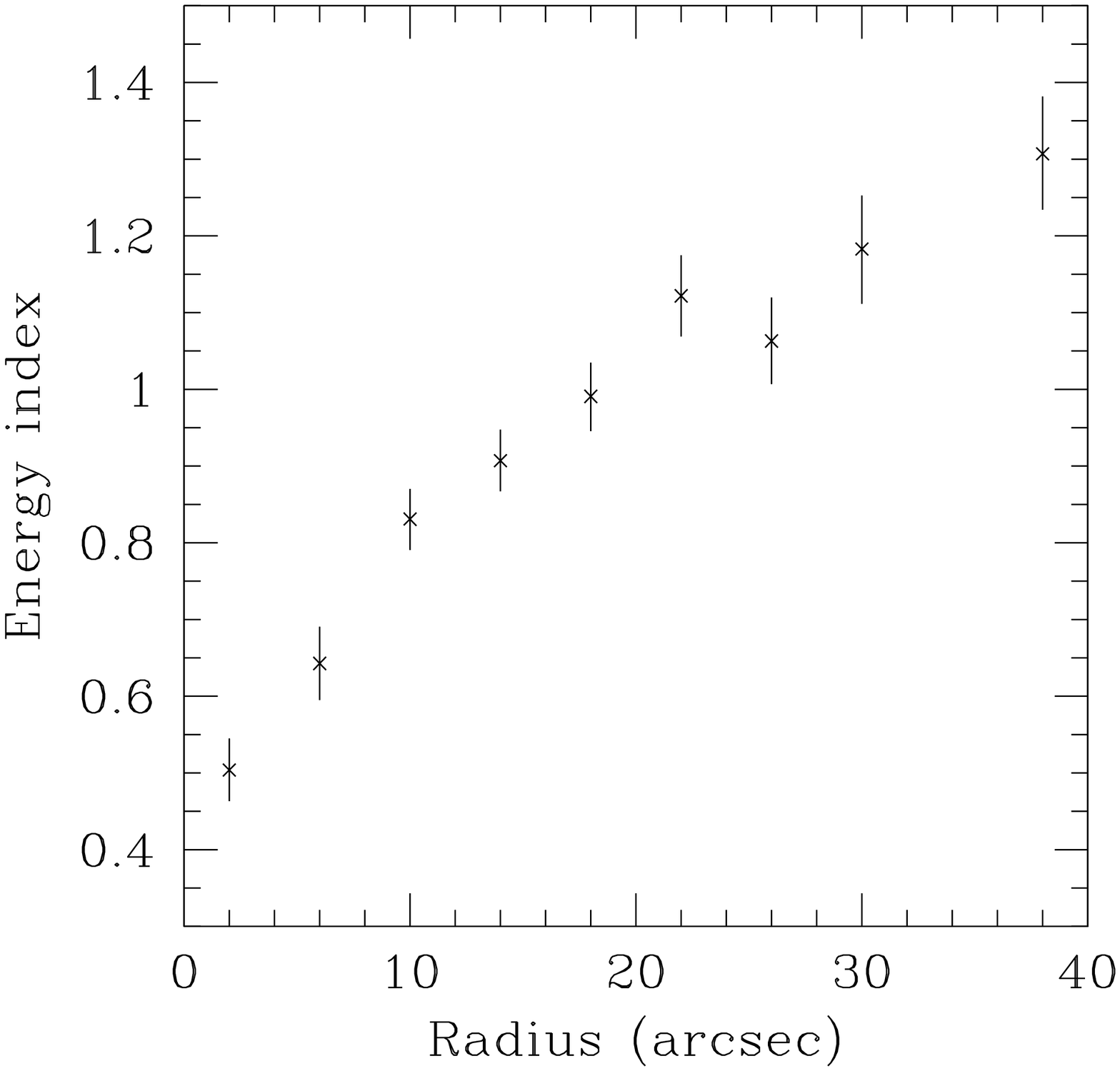,height=2.35in}}
\caption{
Left: {\em Chandra} image of G21.5-0.9. The inset shows the brightness
profile of the compact central region compared with a histogram of
that from a point source in the telescope. Right: Spectral index variation
with radius in the nebula, indicating the synchrotron burn-off of particles
injected at the center of the nebula.
}
\end{figure}

G21.5--0.9 is a compact SNR that exhibits strong linear
polarization, a flat spectrum, and centrally peaked emission in the radio band.
The SNR has a lower $L_x/L_r$ ratio than the Crab; it is a
factor of $\sim 9$ less luminous in the radio and a factor of $\sim 100$
less in X-rays.
To date, there has been no detection of a central pulsar. However,
{\em Chandra}\ observations by Slane et al.~\cite{slane2000}  
reveal a compact source
of emission at the center of the remnant as well as a
radial steepening of the spectral index consistent with synchrotron burn-off
of high energy electrons injected from a central source (Figure~5).
Using an empirical relationship between the total X-ray luminosity
of the PWN with the spin-down power of the pulsar powering the 
nebula~\cite{seward} suggests
the presence of a pulsar with $\dot E = 10^{37.5}{\rm\ erg\ s}^{-1}$, although
Chevalier~\cite{chevalier} 
argues that the spectral variations in the nebula imply
more efficient conversion of $\dot E$ into X-ray emission, and suggests that
$\dot E \approx 10^{36.7}{\rm\ erg\ s}^{-1}$ is more likely for the pulsar in
this PWN. Detection of pulsations from the central source are required to
address this further.
Using the larger $\dot E$ estimate along with pressure estimates from
the radio spectrum, Eq. 3 predicts a wind termination shock at a
radius of $\sim 1.5 \eta^{-1/2}$~arcsec from the pulsar, 
assuming a distance of 5~kpc. As indicated in the inset to Figure~5, 
the brightness profile of the compact X-ray source in G21.5--0.9
is broader than that for a point source. The $\sim 2^{\prime\prime}$ extent
of the source is consistent with the expected size of the 
termination shock zone. The small radius of the termination shock combined
with the large radius of the PWN yields $\sigma \ll 1$; as with the Crab, the
pulsar wind is particle-dominated in the termination shock zone.

Slane et al.~\cite{slane2000} also report the presence of 
a faint extended shell of
emission surrounding G21.5--0.9 whose featureless spectrum may be associated
with energetic particles accelerated by the SNR blast wave, similar to those
observed for SN~1006 and other SNRs as described in Section 2. Safi-Harb 
et al.~\cite{safi} and Warwick et al.~\cite{warwick} confirm the presence
of the shell, but suggest that it is a faint extension of the PWN,
though no extended radio shell is observed.

\subsection{3C58}

3C58 (Figure 6, left) is a young Crab-like supernova remnant. Historical 
evidence strongly
suggests an association of the remnant with supernova SN~1181, which
would make 3C58 younger than the Crab Nebula. Recent \chandra\ observations
have identified the young 65~ms pulsar J0205+6449 at its center,~\cite{murray}
embedded in a compact nebula which appears to be 
confined by the pulsar wind termination shock.~\cite{slane2002}
The central region of 3C58 is shown in Figure 5 (center). 
The emission is clearly extended, with elongation in the N-S direction, 
perpendicular to the long axis of the main nebula; the full extent in this
direction is $\sim 25^{\prime\prime}$. PSR J0205+6449  resides
at the center of this compact nebula. 

\begin{figure}[t]
\centerline{\psfig{file=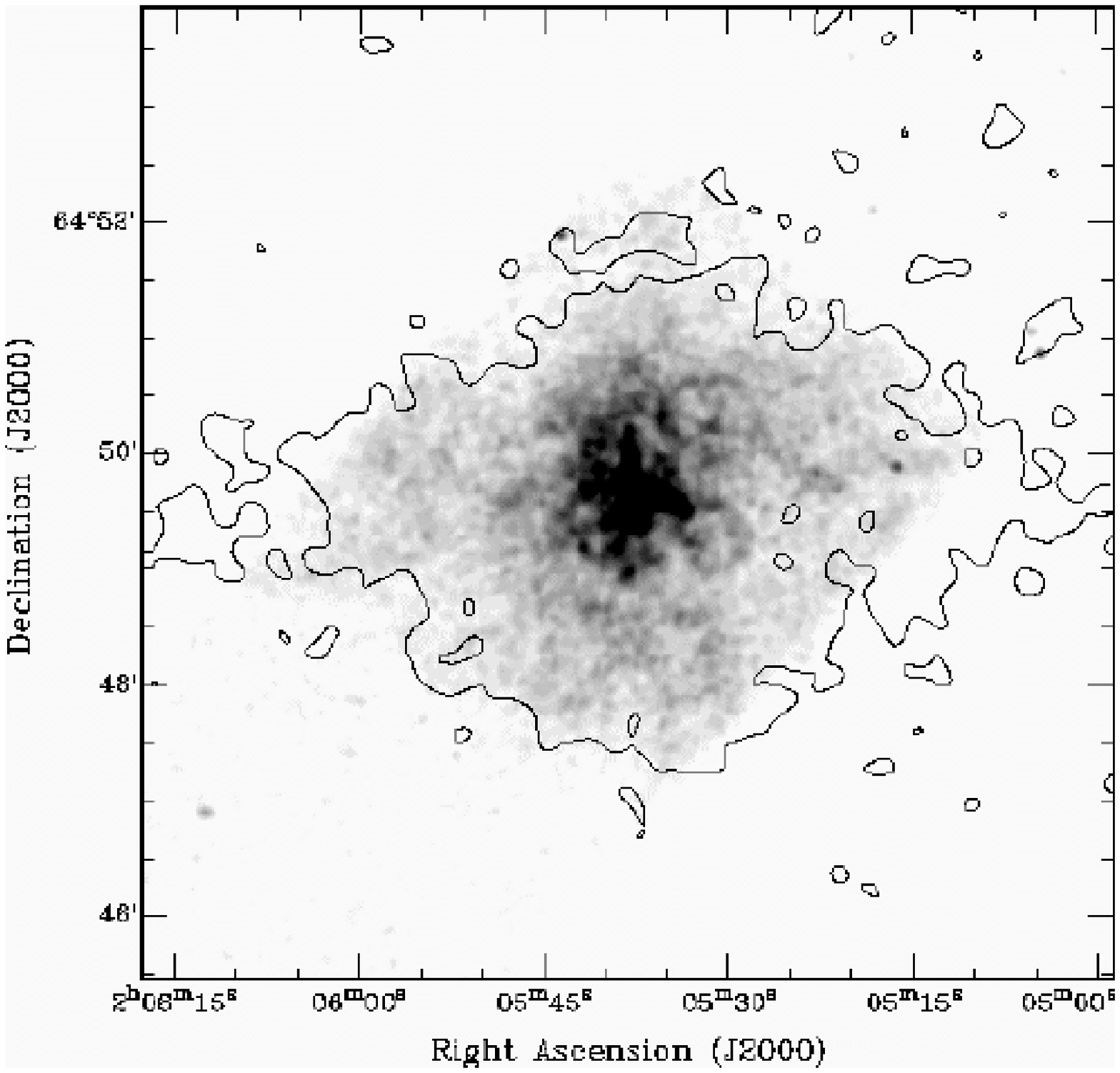,height=2.5in}\psfig{file=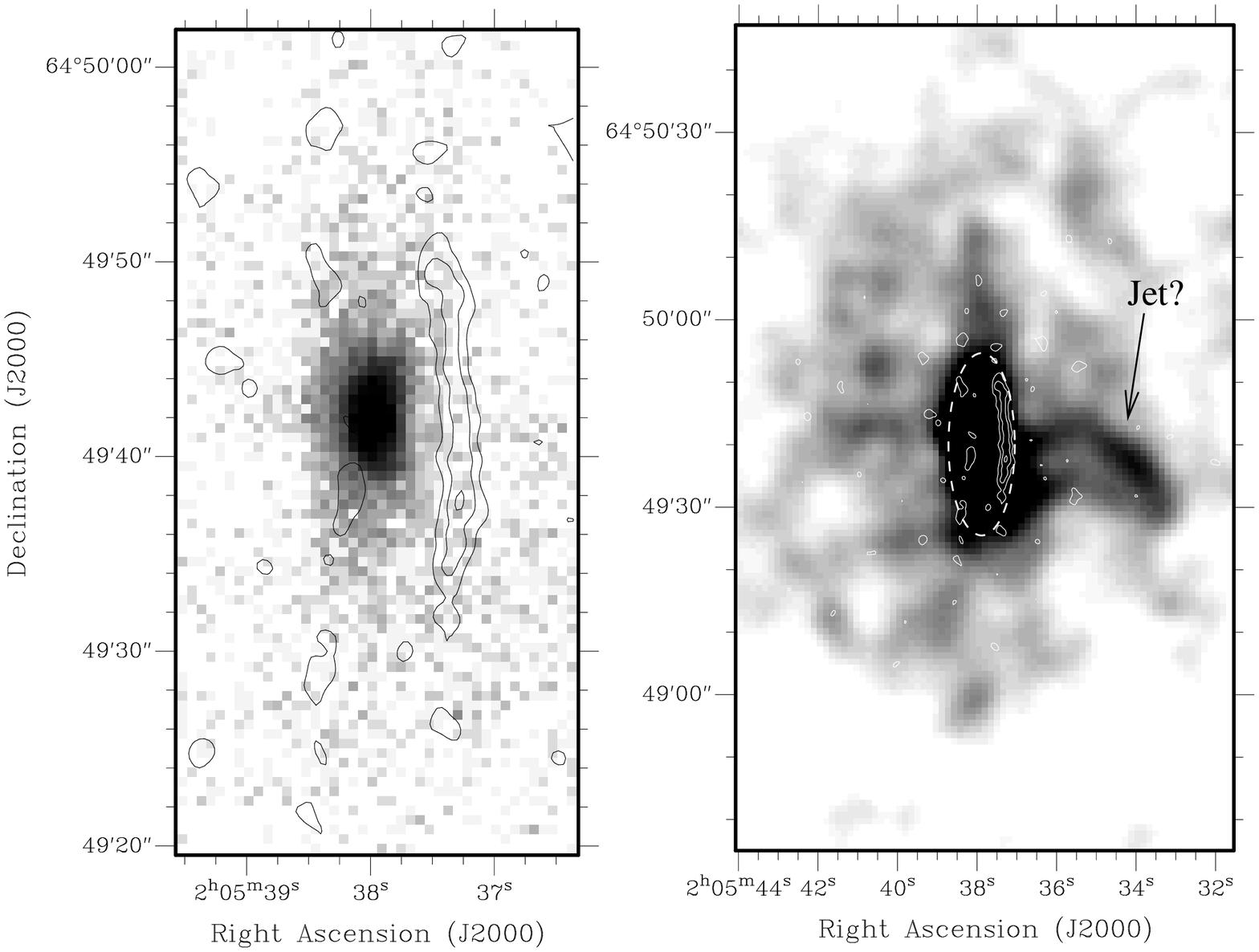,height=2.5in}}
\caption{
Left: ACIS image of 3C 58. Some portions of the nebula extend beyond the
detector boundary, as indicated by the outermost contour from the HRC image, 
which is superposed. Center: ACIS image of the central region of 3C 58,
with contours from 20 cm VLA data showing a faint radio wisp that
bounds the X-ray core. Right: Saturated image of 3C 58 core, on a larger 
scale, revealing the faint jetlike feature extending toward the west. The 
dashed ellipse indicates the rough outline of the extended X-ray core.
}
\end{figure}

The western edge of the compact nebula lies directly along a radio
filament,~\cite{frail} shown as contours in Figure 5 (center),
and there appears to be a slight flattening of compact nebula along this
side. Frail \& Moffett~\cite{frail} suggested that this filament may represent
the position of the termination shock where the pulsar wind is confined
by the interior pressure of the PWN. Adopting the generally accepted distance 
$d =3.2d_{3.2}$~kpc, integration of the radio
synchrotron spectrum yields a pressure of
$P_{\rm neb} = 3.2 \times 10^{-10} d_{3.2}^{-1}{\rm\ dyne\ cm}^{-2}$
under the assumption of equipartition between the electron and magnetic energy
densities. The spin-down properties of the pulsar
give $\dot E = 2.6 \times 10^{37} I_{45} {\rm\ erg\ s}^{-1}$ 
where $I_{45}$ is the NS moment of inertia
in units of $10^{45}{\rm\ gm\ cm}^{2}$. Ram pressure balance should thus
occur at
$r_w = 5.5 \times 10^{17} I_{45}^{1/2} \eta^{-1/2} d_{3.2}^{1/2}$~cm,
or at an angular distance
$\theta = 11.4 I_{45}^{1/2} \eta^{-1/2} d_{3.2}^{-1/2}$~arcsec.
This is in good agreement with the the $\sim 12$~arcsec radial extent of the
core X-ray emission in the NS direction, suggesting that the compact nebula
surrounding the pulsar is bounded by the pulsar wind termination shock. It is
possible that this is actually a toroidal structure, much like
that seen in the Crab Nebula, and that the elongated surface brightness
distribution is the result of
the inclination angle. In this interpretation, the axis of the toroid,
which presumably lies along the rotation axis of the pulsar, lies in the
east-west direction when projected onto the sky. We note that the long
axis of 3C58 itself, as well as an extended jet-like feature shown in 
Figure 5 (right), are both aligned in this direction.
Assuming that the radio filament lies along one side of the torus, its
separation from the pulsar ($\sim 4.5$~arcsec) implies an inclination
angle of $\sim 70^\circ$.

\section{Summary}

The strong shocks in SNRs and pulsar wind nebulae have long been regarded
as sites where particles are accelerated to extremely high energies. SNRs,
in particular, are leading candidates as the source of cosmic rays up to
the knee of the spectrum. X-ray emission from PWNe also require particles
with higher energies than expected in the postshock flow of the pulsar 
wind, indicating efficient acceleration at the termination shock. X-ray
observations have now begun to provide direct evidence of the energetic
particles and shock structures where this such acceleration takes place.
Through studies of the dynamics and nonthermal X-ray emission from the
shells of SNRs, and of the location and structure of the wind termination
shocks and associated particle outflows in PWNe, strong constraints are being
placed on models of the particle acceleration process. High resolution
X-ray observations promise continued advances in this area, through
measurements of SNR nonthermal emission and expansion rates, and the
inner structure of PWNe. In addition, $\gamma$-ray observations with
current and upcoming \v{C}erenkov telescopes, as well as future
space-borne observatories, hold considerable promise for probing these
sites of extremely energetic particles.

\section*{Acknowledgments}
The author would like to thank Bryan Gaensler, Jack Hughes, and Don Ellison
for their contributions as collaborators on much of the above work. This 
research was funded in part by NASA Contract NAS8-39073 and Grants 
NAG5-9281 and GO0-1117A.

\end{document}